\documentclass[12pt]{iopart}
\usepackage{iopams}
\usepackage{textcomp}
\usepackage{graphicx}
\usepackage{exscale}
\usepackage[sort,compress]{cite}
\def\erf{\mathop{\mathrm{erf}}}
\begin{document}

\title[Self-consistent ACF for bias-corrected roughness measurement]{Self-consistent autocorrelation for finite-area bias correction in roughness measurement}

\author{D Ne\v{c}as}
\address{CEITEC, Brno University of Technology, Purky\v{n}ova 123, 61200 Brno, Czech Republic}
\eads{\mailto{yeti@gwyddion.net}}

\begin{abstract}
Scan line levelling, a ubiquitous and often necessary step in AFM data
processing, can cause a severe bias on measured roughness parameters such as
mean square roughness or correlation length. Although bias estimates have
been formulated, they aimed mainly at assessing the severity of the problem
for individual measurements. Practical bias correction methods are still
missing. This work exploits the observation that the bias of autocorrelation
function (ACF) can be expressed in terms of the function itself, permitting
a self-consistent formulation. From this two correction approaches are
developed, both with the aim to obtain convenient formulae which can be easily
applied in practice. The first modifies standard analytical models of ACF to
incorporate, in expectation, the bias and thus actually match the data the
models are used to fit. The second inverts the relation between true and
estimated ACF to realise a model-free correction. Both are tested using
simulated and experimental data and found effective.
\end{abstract}

\vspace{2pc}

\section{Introduction}
\label{sec:intro}

Recently a couple of works drew attention to how roughness measurement by
atomic force microscopy (AFM) are impacted by levelling/background
subtraction~\cite{Necas20,Necas20_SR}, in particular line levelling,
a ubiquitous and often necessary step in AFM data
processing~\cite{PascualStarink1996,Erickson2012,Wang2018,Marinello2007}. The
classic results for the effect of mean value subtraction on statistical
quantities~\cite{Anderson71,Zhao00,Krishnan15} were generalised in
a theoretical framework covering many common levelling methods. The mean
square roughness $\sigma$, as well as many other quantities, becomes biased.
For 1D data and 1D scan line levelling the bias of estimate $\hat\sigma$ can
be written (in expectation):
\begin{equation}
\mathrm{E}[\hat\sigma^2] = \sigma^2 - 2 \int_0^1 G(tL)\,C(t)\,\rmd t\;.
\label{bias-sigma2}
\end{equation}
Function $G$ is the true autocorrelation function (ACF) of the roughness and
$C$ a complicated function capturing correlation/spectral properties of the
specific levelling method. The second term expresses the measurement bias,
which can often be severe~\cite{Zhao00,Necas20,Necas20_SR}.
Explicit expressions are known for several common levelling methods and
autocorrelation function forms.~\cite{Necas20} It should be noted that is more
correct to call $G$ the autocovariance function and reserve the term
autocorrelation for the function normalised to variance, but both are commonly
used. The bias problem is not unique to AFM and profilometry data levelling.
Similar problems occur for autocovariance function estimation from locally
smoothed (detrended) data~\cite{Hyndman1997,Park2009}.

Ultimately, the bias and variance depend on the ratio $\alpha=T/L$ of
correlation length $T$ and scan line length $L$. The bias further increases
with `aggressivity' of the levelling procedure~\cite{Zhao00,Necas20}. The
ratio $\alpha$ must be kept small for reliable results. If scan line levelling
and similar 1D corrections are applied to images the error is proportional to
$\alpha$ (not $\alpha^2$ as one might assume for 2D image data), which can be
difficult to keep sufficiently small. Even when scan lines are not levelled
explicitly, the computation of 1D ACF imposes the condition of zero mean value
on image rows, corresponding to degree-0 polynomial levelling. The length $L$
is, sadly, also often not set deliberately but instead to what `feels
right'~\cite{Necas20_SR}. This then translates to $\alpha$ which is way too
large---sometimes far beyond instrumental constraints. Reported results are
then unnecessarily skewed. Bias estimation procedures have been proposed,
either simple and coarse~\cite{Necas20_SR} or more detailed~\cite{Necas20},
allowing one to check whether it is within reasonable bounds. The simplest
(and coarsest) estimate of relative bias of $\hat\sigma$ is $-n\alpha$, where
$n$ is the number of terms in the scan line levelling polynomial, usually
equal to its degree plus one.

\begin{figure}
\includegraphics[width=0.5\hsize]{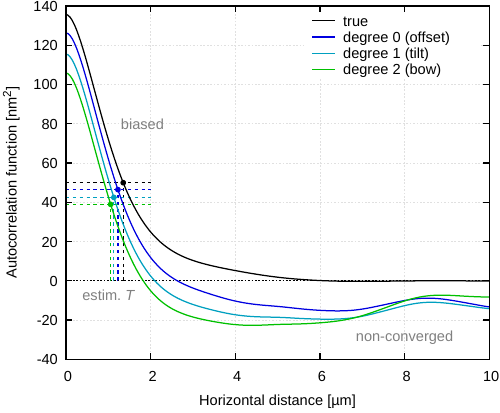}
\caption{The effect of scan line polynomial levelling using polynomials of
various degrees on the estimated ACF. The beginning of the curve (small
distances) is biased, whereas for larger distances the ACF estimate is not
converged and exhibits oscillations~\cite{Zhao00}.}
\label{fig:ACF-motivation}
\end{figure}

Unfortunately, all the estimates suffer from a chicken and egg problem.
They require knowing the correlation length $T$, or even the form of the ACF,
which are not known {\it a priori}. They should be the outputs of our
measurement. Therefore, they must be estimated from experimental data, and
these estimates are again biased. The experimental $T$ (denoted $\hat T$) is
underestimated because the entire ACF is affected in a similar manner as
$\sigma^2$, as illustrated in \fref{fig:ACF-motivation}. Consequently,
although such estimates can help with judging the bias for a particular
measurement or guide towards a better choice of scanning parameters, they
cannot be applied in a logically consistent manner. They are thus of limited
use for actual correction of the biased results. Clearly, the problem is not
yet satisfactorily solved. In order to deal with the bias pervading all the
roughness parameters we need a self-consistent method which does not require
{\it a priori} knowledge of the result. It should also be convenient to have
practical impact and allow wide adoption. Here we aim to provide this missing
piece.

The overall plan is fairly straightforward. We being from the observation that
the value of ACF at zero is $\sigma^2$, that is $\sigma^2=G(0)$. Formula
\eref{bias-sigma2} can thus be also written
\begin{equation}
\mathrm{E}[\hat G(0)] = G(0) - 2 \int_0^1 G(tL)\,C(t)\,\rmd t\;.
\end{equation}
The second (bias) term is linear in $G$. Suppose an expression of the same
form could be obtained for $G$ as a whole (we show later that it is indeed the
case)
\begin{equation}
\mathrm{E}[\hat G] = G - RG\;.
\label{EGhat}
\end{equation}
Here $R$ is a linear operator expressing the bias, now of the entire ACF. It
again captures the properties of the specific levelling procedure. Expression
\eref{EGhat} ties self-consistently together the true and estimated ACF. We
can say that the ACF knows about its own bias. The relation can be formally
inverted
\begin{equation}
G = (1 - R)^{-1}\mathrm{E}[\hat G]\;,
\end{equation}
yielding unbiased $G$ from the biased estimate $\hat G$ (in expectation). This
is the adventurous option---it is not immediately obvious such inversion would
be numerically feasible.

The conservative option is to employ expression \eref{EGhat} directly. Assume,
for instance, that the roughness is Gaussian. The true ACF has then the form
\begin{equation}
G_\mathrm{Gauss}(\tau) = \sigma^2\exp\left(-\frac{\tau^2}{T^2}\right)\;.
\label{GGauss}
\end{equation}
Conventionally, we fit the experimental ACF $\hat G(\tau)$ with the ideal
model $G_\mathrm{Gauss}(\tau)$ with $\sigma$ and $T$ as free parameters. But
it is clearly the wrong model. It does not describe the experimental ACF,
which never conforms to the theoretical form. The correct model is
\begin{equation}
G_\mathrm{Gauss}(\tau)-RG_\mathrm{Gauss}(\tau)
\end{equation}
and can be obtained by applying $R$ to $G_\mathrm{Gauss}(\tau)$.

The questions are what is the form or operator $R$, whether $R$ and
$(1-R)^{-1}$ can be reasonably evaluated and how well the bias correction
works in practice. They are answered in the following sections. The general
expression for $R$ is derived in \sref{sec:biascalc}, which the reader can
skip on the first reading. \Sref{sec:practical} provides elementary formulae
and procedures for practical bias correction and \sref{sec:examples} tests
their effectiveness using simulations and real AFM data.

\section{Bias of ACF after levelling}
\label{sec:biascalc}

The calculation of $R$ follows the general scheme and notation introduced in
Ref.\ \citenum{Necas20} (sections 3.1 and 3.3), including treating the data as
continuous functions. Since scan line levelling is the dominant source of bias
even for image data~\cite{Necas20_SR}, we consider the 1D case. Denote
$\varphi_j$ orthonormal basis functions used for background subtraction by
linear fitting, with $j$ distinguishing the functions. If $\varphi_j$ are
polynomials then $j\in\{0,1,2,\dots n-1\}$ is their degree, but the index may
not be a simple integer in other cases. Summations over $j$ are, therefore,
written below only formally.

Levelled data are computed by subtracting the projection onto the span of
$\varphi_j$
\begin{equation}
\hat z(x) = z(x) - \sum_j a_j \varphi_j(x)\;,
\label{hatz}
\end{equation}
with coefficients $a_j$ equal to the dot products
\begin{equation}
a_j = \int_0^L z(x)\,\varphi_j(x)\,\rmd x\;.
\label{aj}
\end{equation}
The ACF is estimated as
\begin{equation}
\hat G(\tau) = \frac1{L-\tau} \int_0^{L-\tau} \hat z(x)\,\hat z(x+\tau)\,\rmd x\;.
\label{hatGint}
\end{equation}
Substituting expressions \eref{hatz} into \eref{hatGint} gives for $\hat G(\tau)$
\begin{equation}
\frac1{L-\tau} \int_0^{L-\tau}
\left[z(x) - \sum_j a_j \varphi_j(x)\right]
\left[z(x+\tau) - \sum_k a_k \varphi_k(x+\tau)\right]
\rmd x\;,
\end{equation}
which can be expanded into four terms corresponding to the combinations of
$z$ and $\varphi$:
\begin{equation}
\hat G(\tau) = \hat G_1(\tau) - \hat G_2(\tau) - \hat G_3(\tau) + \hat G_4(\tau)\;.
\end{equation}
Taking expectations,
\begin{equation}
\eqalign{
\mathrm{E}[\hat G_1(\tau)]&= G(\tau)\cr
\mathrm{E}[\hat G_2(\tau)]&= \frac1{L-\tau} \sum_j \int_0^{L-\tau} \int_0^L
                             G(x'-x+\tau)\,\varphi_j(x)\,\varphi_j(x')
                             \,\rmd x'\,\rmd x\cr
\mathrm{E}[\hat G_3(\tau)]&= \frac1{L-\tau} \sum_k \int_0^{L-\tau} \int_0^L
                             G(x'-x)\,\varphi_k(x+\tau)\,\varphi_j(x')
                             \,\rmd x'\,\rmd x\cr
\mathrm{E}[\hat G_4(\tau)]&= \frac1{L-\tau} \sum_{j,k} \left[\int_0^L \int_0^L
                             G(x''-x')\,\varphi_k(x')\,\varphi_j(x'')
                             \,\rmd x'\,\rmd x''\right.\cr
                          &\hphantom{=\frac1{L-\tau} \sum[}\times
                             \left. \int_0^{L-\tau}\varphi_j(x)\varphi_k(x+\tau)
                             \,\rmd x\right]\;,
}
\label{EG1234monster}
\end{equation}
where we utilised the linearity of expectation and that for any $a$ and $b$
\begin{equation}
\mathrm{E}[z(a)z(b)] = G(b-a) = G(a-b)\;.
\end{equation}
In a similar manner as in formula \eref{bias-sigma2} for the bias of
$\sigma^2$, one term (here $\mathrm{E}[\hat G_1]$) gives the unbiased
$G(\tau)$ and the remaining terms combine to give the bias $RG(\tau)$.

\subsection{Linear operator $R$}

In principle, formulae \eref{EG1234monster} can already be
considered a representation of the operator $R$. However, it is more natural
(and useful) to write it explicitly
\begin{equation}
\mathrm{E}[\hat G(\tau)] = G(\tau) - \int_0^L R(\tau,u)\,G(u)\,\rmd u\;.
\label{EGintRG}
\end{equation}
Meaning in $\mathrm{E}[\hat G_2(\tau)]$ we must set $u=x'-x+\tau$, transform
the domain of integration (which splits the integral into three) and obtain
\begin{equation}
\eqalign{
\mathrm{E}[\hat G_2(\tau)] = \frac1{L-\tau} \sum_j \Bigl[\Bigr.
        & \int_0^{L-\tau} G(u)\,c_{j,[0,L-\tau-u]}(\tau+u)\,\rmd u\cr
        &+ \int_0^\tau G(u)\,c_{j,[0,L-\tau]}(\tau-u)\,\rmd u\cr
        &+ \int_\tau^L G(u)\,c_{j,[u-\tau,L-\tau]}(\tau-u)\,\rmd u
           \Bigl.\Bigr]\;.
}
\end{equation}
The symmetry of $G$ was utilised to ensure its argument is always positive and
thus from interval $[0,L]$. Functions $c_j$ again express the correlation
properties of $\varphi_j$, in analogy to Ref. \citenum{Necas20}. However, as
the various integrals are over different subintervals of $[0,L]$ they are more
complicated here, defined
\begin{equation}
c_{j,[a,b]}(\delta) = \int_a^b \varphi_j(x)\,\varphi_j(x+\delta)\,\rmd x\;.
\label{cj}
\end{equation}
Finally, in order to transform the expression to the form \eref{EGintRG}, we
replace the integration limits for $u$ using the indicator function
\begin{equation}
\chi_{[a,b]}(x) = \cases{1&if $a\le x\le b$\cr0&otherwise}
\end{equation}
resulting in
\begin{equation}
\eqalign{
\mathrm{E}[\hat G_2(\tau)] = \frac1{L-\tau} \int_0^L G(u) \sum_j \Bigl[\Bigr.
        & \chi_{[0,L-\tau]}(u)\,c_{j,[0,L-\tau-u]}(\tau+u)\cr
        &+ \chi_{[0,\tau]}(u)\,c_{j,[0,L-\tau]}(\tau-u)\cr
        &+ \chi_{[\tau,L]}(u)\,c_{j,[u-\tau,L-\tau]}(\tau-u)
           \Bigl.\Bigr]\,\rmd u\;.
}
\end{equation}
The term in square brackets is one piece of $R(\tau,u)$ in the form required
by \eref{EGintRG}---the one corresponding to $\hat G_2$. The second piece,
corresponding to $\hat G_3$, is obtained using the same steps. The last piece
contains integrals combining $\varphi_j$ and $\varphi_k$ for $j\ne k$ that
cannot be expressed using \eref{cj}. If we define
\begin{equation}
c_{j,k,b}(\delta) = \int_0^b \varphi_j(x)\,\varphi_k(x+\delta)\,\rmd x\;.
\label{cjk}
\end{equation}
it can be written
\begin{equation}
\mathrm{E}[\hat G_4(\tau)]
= \frac1{L-\tau} \int_0^L G(u) \sum_{j,k}
  [c_{j,k,L-u}(u) + c_{k,j,L-u}(u)]c_{j,k,L-\tau}(\tau)] \,\rmd u
\end{equation}
Therefore, the final expression for $R(\tau,u)$ is
\begin{equation}
\eqalign{
R(\tau,u) &= \frac1{L-\tau} \sum_j \Bigl[
   \chi_{[0,L-\tau]}(u) \bigl(c_{j,[0,L-\tau-u]}(\tau+u) + c_{j,[\tau+u,L]}(-\tau-u)\bigr)
   \Bigr.\cr
   &\hphantom{=\frac1{L-\tau}\sum}
   + \chi_{[0,\tau]}(u) \bigl(c_{j,[0,L-\tau]}(\tau-u) + c_{j,[\tau,L]}(u-\tau)\bigr)\cr
   &\hphantom{=\frac1{L-\tau}\sum}
   + \chi_{[\tau,L]}(u) \bigl(c_{j,[u-\tau,L-\tau]}(\tau-u) + c_{j,[\tau,L+\tau-u]}(u-\tau)\bigr)
           \Bigl.\Bigr]\cr
   &\hphantom{=} - \frac1{L-\tau} \sum_{j,k}
    \bigl[c_{j,k,L-u}(u) + c_{k,j,L-u}(u)\bigr]c_{j,k,L-\tau}(\tau)
}
\label{Rtauu}
\end{equation}

\subsection{Polynomial levelling}
\label{sec:poly}

\begin{figure}
\includegraphics[width=0.5\hsize]{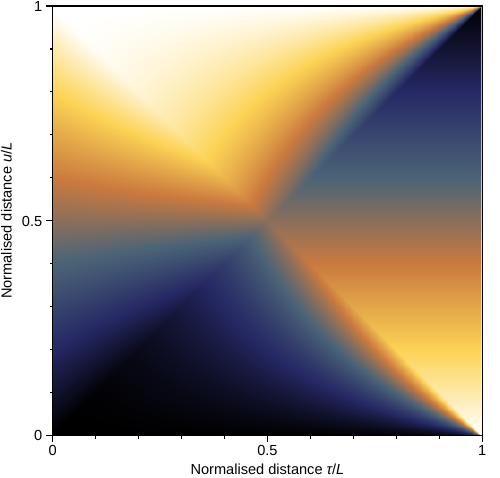}
\caption{Linear operator $R_1(\tau,u)$ for mean value subtraction, normalised
to $L$ (equivalent to putting $L=1$). The false colour scale corresponds to
the interval $[0,2]$.}
\label{fig:R0-map}
\end{figure}

A polynomial basis $\varphi_j$ has symmetries which can be used simplify
$R(\tau,u)$ somewhat. We first note that the expression is not unwieldy
because we failed to express it more elegantly. The operator is inherently
complicated, with a number of discontinuities in the derivative. Even for mean
value subtraction, when the single basis function $\varphi_0$ is a constant,
we get
\begin{equation}
R_1(\tau,u) = \frac2L \left[\chi_{[0,L-\tau]}(u)\frac{L-\tau-u}{L-\tau}
                            + \chi_{[0,\tau]}(u)\frac{u-\tau}{L-\tau}
                            + \frac{(L-u)\tau}{L(L-\tau)}\right]\;,
\end{equation}
illustrated in \fref{fig:R0-map}. Although only function values for small
$\tau$ and $u$ are important and some of the discontinuities do not affect
expansions for small $\tau$ and $u$, $R(\tau,u)$ is not totally differentiable
at $(0,0)$. A small-$\tau$ approximation of the entire integral in
\eref{EGintRG} is possible only because the integral is a smoother function
than $R(\tau,u)$ itself.

For general polynomials, we note that Legendre polynomials $\mathrm{P}_n(x)$
on interval $[-1,1]$ are either even or odd,
$\mathrm{P}_n(-x)=(-1)^n\mathrm{P}_n(x)$. For the orthonormal basis functions
on $[0,L]$ it translates to
\begin{equation}
\varphi_j(L-x) = (-1)^j \varphi(x)\;.
\end{equation}
From this we can easily see that
\begin{equation}
c_{j,k,L-\tau}(\tau) = (-1)^{j+k} c_{k,j,L-\tau}(\tau)\;.
\end{equation}
Terms with $j+k$ odd can be omitted as they mutually cancel. And for $j+k$
even only terms with $j<k$ can be kept, multiplied by 2. Together with the
relation $c_{j,j,b}=c_{j,[0,b]}$, permitting rewriting terms with $j=k$, these
rules eliminate most of the terms in the second summation in \eref{Rtauu}. In
fact, for degree 1 no such term remains, giving
\begin{equation}
\eqalign{
R_2(\tau,u) &= \frac1{L-\tau} \sum_{j=0}^1 \Bigl[
   \chi_{[0,L-\tau]}(u) \bigl(c_{j,[0,L-\tau-u]}(\tau+u) + c_{j,[\tau+u,L]}(-\tau-u)\bigr)
   \Bigr.\cr
   &\hphantom{=\frac1{L-\tau}\sum}
   + \chi_{[0,\tau]}(u) \bigl(c_{j,[0,L-\tau]}(\tau-u) + c_{j,[\tau,L]}(u-\tau)\bigr)\cr
   &\hphantom{=\frac1{L-\tau}\sum}
   + \chi_{[\tau,L]}(u) \bigl(c_{j,[u-\tau,L-\tau]}(\tau-u) + c_{j,[\tau,L+\tau-u]}(u-\tau)\bigr)\cr
   &\hphantom{=\frac1{L-\tau}\sum}
   - 2c_{j,[0,L-u]}(u)c_{j,[0,L-\tau]}(\tau) \Bigl.\Bigr]\cr
}
\end{equation}
A similar simplification is possible for other bases formed by even and odd
functions $\varphi_j$, for instance sines and cosines, although the indexing
by $j$ may differ (and sines and cosines are more natural to handle in the
frequency domain). However, the small-$(\tau,u)$ expansion for a specific
basis is still tedious and better evaluated using symbolic algebra software.

Maxima~\cite{Maxima} was used to obtain the practical formulae summarised in
the following section. The expansions were terminated at $\alpha^2$ terms.
The first reason is that preliminary numerical experiments showed that the
leading $\alpha^1$ terms is not always sufficient and without the second term
there is a tendency to overcorrection. The general form of $\sigma^2$ bias for
polynomial levelling contains only even-power terms after $\alpha^2$ (equation
(27) in Ref.\ \citenum{Necas20}). Therefore, there is no third order term in
the expansion and higher powers are negligible. Finally, the low smoothness of
$R$ at zero means that more accurate expansions would not be, in general,
Taylor-like and would have to include more complicated ACF-specific terms. For
this reason it is advantageous to express analytical models of ACF in terms of
$\alpha=T/L$ and $s=\tau/T$ as it makes them smoother functions. In model-free
inversion there is no $T$. Therefore, the formulation has to be done in terms
of $t=\tau/L$ instead of $s$.

For a polynomial with $n$ terms (degree $n-1$) the expansion up to the
second order in $\alpha$ is
\begin{equation}
R_n G(\tau) = \frac{2n}{L-\tau}\left[ M_0
              - \frac{n(L+\tau)}{L^2} M_1
              - \frac{n}{L} I(\tau) \right]\;,
\label{RnG}
\end{equation}
where
\begin{equation}
M_0 = \int_0^\infty G(u)\,\rmd u \;,\quad
M_1 = \int_0^\infty uG(u)\,\rmd u
\label{M0M1}
\end{equation}
and
\begin{equation}
I(\tau) = \int_0^\tau (\tau-u)G(u)\,\rmd u\;.
\label{Itau}
\end{equation}
The expressions in the following section are obtained by evaluating
\eref{RnG} for particular $G$.

\section{Practical formulae}
\label{sec:practical}

\subsection{Corrected models}
\label{sec:biased-model}

The (biased) discrete ACF is estimated from data values $z_k$~\cite{Zhao00}
\begin{equation}
\hat G_k = \frac1{N-k}\sum_{m=0}^{N-1-k} z_m z_{m+k}\;,
\label{Gdiscrete}
\end{equation}
where $\tau=k\Delta_x$ if $\Delta_x$ is the sampling step. It is fitted by an
ACF model function. Simple models have only two free parameters, $\sigma$
and~$T$. The classic Gaussian ACF model \eref{GGauss} and analogous
exponential model
\begin{equation}
G_\mathrm{exp}(\tau) = \sigma^2\exp\left(-\frac{\tau}T\right)
                     = \sigma^2\rme^{-s}\;.
\label{Gexp}
\end{equation}
are replaced with the leading terms of $(G-BG)(\tau)$ expanded for small
$\alpha$ and $\tau$. In particular, the Gaussian model is replaced with
\begin{equation}
G_\mathrm{Gauss}^\mathrm{bias}(\tau)
= \sigma^2\left[\rme^{-s^2}(1+n^2\alpha^2)
               - \sqrt{\pi}n\alpha
               + \sqrt{\pi}n\alpha^2s \bigl(n\erf(s) - 1\bigr)\right]
\label{GGauss-biased}
\end{equation}
and the exponential model with
\begin{equation}
G_\mathrm{exp}^\mathrm{bias}(\tau)
= \sigma^2\left[\rme^{-s}(1 + 2n^2\alpha^2)
                - 2n\alpha
                + 2n(n-1)\alpha^2s\right]\;.
\label{Gexp-biased}
\end{equation}
where $s=\tau/T$, $\alpha=T/L$ and erf denotes the error function
(antiderivative of Gaussian). If evaluation of special functions is not
possible or desirable erf can be replaced for instance by a Pad\'e-style
approximation as it only occurs in the second order term. The
intermediate Gaussian--exponential ACF model~\cite{Franceschetti07,Zhao00,Necas20}
\begin{equation}
G_p(\tau) = \sigma^2 \exp\left[-\left(\frac{\tau}T\right)^p\right] = \sigma^2 \exp(-s^p)
\label{Ggenp}
\end{equation}
is replaced with
\begin{equation}
\eqalign{
G^\mathrm{bias}_p(x)
        &= \sigma^2 \left[ \exp(-s^p)
                          - n\alpha \Gamma\left(\frac1p\right)(1+\alpha s)
                          \right. \cr
        &\left.\hphantom{\sigma^2\;\Bigl[\Bigr.]}
            + n^2\alpha^2\Gamma\left(\frac2p\right)
            + n^2\alpha^2s\gamma\left(\frac1p, s^p\right)
            - n^2\alpha^2\gamma\left(\frac2p, s^p\right)  \right]\;, \cr
}
\label{Ggenp-biased}
\end{equation}
where $\Gamma$ denotes the gamma function and $\gamma$ the lower
incomplete gamma function.

The superscript $^\mathrm{bias}$ indicates the models are bias-corrected,
i.e. take into account the
bias of the data \eref{Gdiscrete} they are used to fit. Models
\eref{GGauss-biased} and \eref{Gexp-biased} should be fitted from zero to
approximately the first zero crossing, i.e.\ up to the first $k$ for which
$G_k<0$. The biased models do not have any additional free parameters.
Nevertheless, they contain two additional inputs, the profile or scan line
length $L$ and the number of terms $n$ of the line levelling
polynomial---which is one plus its degree. The full profile length must be
entered as $L$, not the length of ACF data which are often cut to a shorter
interval of $\tau$.

\subsection{Model-free inversion}
\label{sec:blind}

For an unknown, but quickly decaying ACF, formulae
\eref{RnG}--\eref{Itau} can be evaluated using the discrete values of
estimated $\hat G_k$, leading to the following expressions:
\begin{equation}
\hat G_k = \sum_{j=0}^{K-1} A_{kj} G^\mathrm{c}_j\;,
\label{GAGc}
\end{equation}
where $G^\mathrm{c}_m$ with $m=0,1,2,\dots,K-1$ are the correct ACF values
and matrix $A$ is
\begin{equation}
A_{kj} = \delta_{k,k} - \frac{2n}{N-k} + \frac{2n^2}{N^2} \frac{N+k}{N-k}j
         + \frac{2n^2}{N} \frac{k-j}{N-k} \chi(j<k)\;.
\label{Akm}
\end{equation}
Although \eref{GAGc} can be read as expressing measured $\hat G_k$ using
a true ACF $G^\mathrm{c}_m$, we interpret it as a set of $K$ linear equations
for corrected ACF $G^\mathrm{c}_m$, with $A$ being matrix of the system.
Number $K$ is the cut-off after which the function is assumed to be negligible
or the data not usable, i.e.\ again around the first zero crossing.
Symbol $\chi(j<m)$ is 1 when $j<m$ and 0 otherwise.

Matrix $A$ is the sum of four simple matrices, the identity matrix, two
rank-1 matrices and a lower diagonal matrix (in this order). The most
efficient solution may be solving first a system with only the first and last
terms
\begin{equation}
A_{kj}' = \delta_{k,j} + \frac{2n^2}{N} \frac{k-j}{N-k} \chi(j<k)\;,
\end{equation}
as $A'$ is lower triangular and thus the equations are solved by
back substitution. Sherman--Morrison or Woodbury
formula~\cite{Guttman1946,Bartlett1951} is then used to
perform low-rank updates of the solution to include the two rank-1 terms.
However, numerical stability of the update formulae is not well understood.
Furthermore, the full system is well-conditioned and only moderately sized.
Therefore, it can be easily solved using any standard linear algebra routine.

\section{Numerical and experimental examples}
\label{sec:examples}

\subsection{Simulated data---Gaussian ACF}

We first compare the performance of standard and biased Gaussian
roughness models \eref{GGauss} and \eref{GGauss-biased} using simulated
data. Synthetic rough Gaussian surfaces were generated using the spectral
method with $T=20\,\hbox{px}$. The correlation length is in the typical range
for real AFM images, regardless of the physical dimensions of the scanned
area. The mean square roughness $\sigma$ was set to 1 as it is only a scaling
parameter. The image size varied from 100\,px to 2000\,px, corresponding to
$\alpha$ from 0.01 to 0.2 (in the reverse order). The discrete ACF
\eref{Gdiscrete} was evaluated using the standard Fast Fourier Transform
method, after levelling image rows using polynomials with degrees from 0 to 2.

The polynomial levelling was, of course, not actually necessary here because
the simulated data were ideal and had neither tilt nor bow. It simulated the
effect of preprocessing that would be applied to measured data. Tilt or bow
could be added beforehand, but it would be pointless. The levelling would
subtract them again, together with a part of the roughness---which is the
effect we are studying.

\begin{figure}
\includegraphics[width=\hsize]{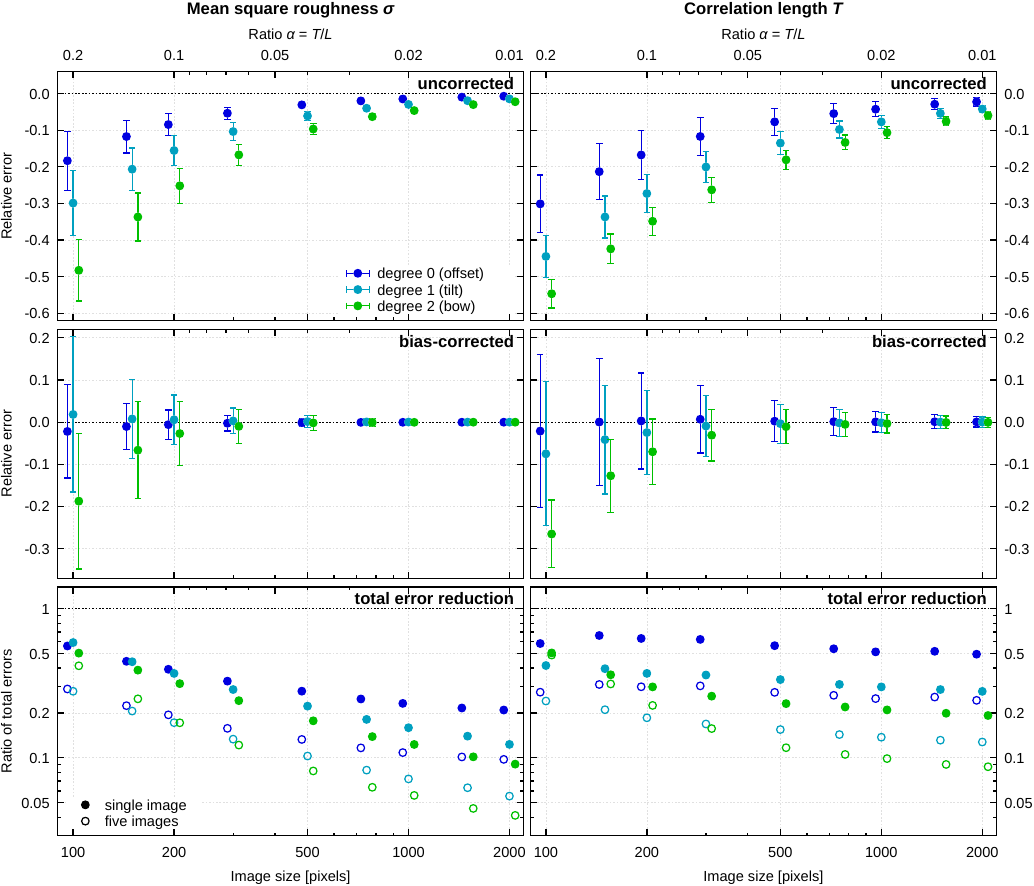}
\caption{Comparison of fitting the biased Gaussian ACF model
\eref{GGauss-biased} with the standard one \eref{GGauss} (for correlation
length of 20\,px).
Error bars represent single-image standard deviations.
Results for different polynomial degrees are slightly offset horizontally for
visual clarity.}
\label{fig:corrected-model-fit}
\end{figure}

Marquardt--Levenberg algorithm was used for the non-linear least squares
fitting to obtain $\sigma$ and~$T$. Both models were fitted to data up to the
first zero crossing. The entire procedure was repeated with randomly generated
Gaussian surfaces hundreds of times (with more repetitions for smaller images
for which the variances are larger). The means and standard deviations are
plotted in \fref{fig:corrected-model-fit}.

The biased model \eref{GGauss-biased} clearly succeeded at bias reduction. For
both parameters and almost all image sizes the bias becomes so small that it
is no longer an issue. The only exception is very small images which are only
several correlation lengths large ($\alpha\lesssim1/10$). Although bias
usually still decreases, it is at the cost of considerably increased variance.
Too much roughness information is missing in such small areas. Using them for
roughness evaluation is just wrong and the correction cannot change it.

The correction generally trades the bias for variance, i.e.\ the parameters
have larger variances than for the standard model. For reasonable $T/L$
the trade-off is advantageous. The total error
$(\hbox{variance}+\hbox{bias}^2)^{1/2}$ decreases as illustrated in the
bottom row of \fref{fig:corrected-model-fit}. The improvement is more marked
for $\sigma$ where it can be an order of magnitude, whereas for $T$ it ranges
from about $2\times$ to $5\times$. The improvement is larger for higher
polynomial degrees. It is because the bias is larger in absolute value, but of
the same functional form. Hence, the same correction is able to deal with
a larger bias.

Full circles in \fref{fig:corrected-model-fit} correspond to the worst case
scenario of a single-image roughness measurement. Multiple scans reduce the
variance, the dominant contribution for the improved model, but do not help
with bias, the dominant contribution for the standard one. This is illustrated
in the plot of total error reduction for five-image evaluation (open circles).

\subsection{Simulated data---inversion}
\label{sec:pyramids}

\begin{figure}
\includegraphics[width=\hsize]{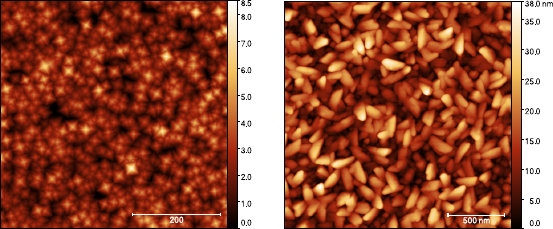}
\caption{Small parts ($512\times512$) of surfaces used in the model-free
correction (inversion) examples. Left: simulated random pyramids. Right:
polycrystalline SrO thin film.}
\label{fig:surfaces}
\end{figure}

For model-free correction (inversion) random pyramidal surfaces with an unknown ACF
were generated using Gwyddion~\cite{Necas12} {\it Objects} function which
generates surfaces by sequential `extrusion'~\cite{Necas21}. The pyramids were
randomly oriented and the pattern was large-scale isotropic. The generated
images were $8000\times8000$ pixels, corresponding to approximately 700
correlation lengths $(\alpha\approx0.0015)$. A small ($512\times512$) part of
one such image is shown in \fref{fig:surfaces}. Smaller images of various
sizes were again cut from the large base image and used to estimate the ACF.

The corrected ACF was computed by cutting $\hat G_k$ slightly beyond the first
zero crossing (10\,\% farther) and solving the linear system
\eref{GAGc} as described in \sref{sec:blind}. Roughness parameters
$\sigma$ and $T$ were again evaluated from both the biased and corrected ACF.
In particular $\sigma$ was calculated from the relation $\sigma^2=G_0$
and $T$ as the distance at which the ACF first falls to $G_0/\rme$ ($\rme$
being Euler's number). The $1/\rme\approx0.368$ threshold is consistent with
the analytic models \eref{GGauss} and \eref{Gexp}, although it should be noted
that roughness measurement standards often set the threshold differently, 0.2
being a common choice~\cite{ISO-25178}.

The comparison also requires true values of $\sigma$ and~$T$. They were
obtained using angularly averaged 2D ACF, which was averaged over all
generated images. The data were artificial and did not contain any tilt, bow,
sample bending or other type of background. Therefore, the only preprocessing
necessary before the computation of 2D ACF was the subtraction of the mean
value from the entire image. The relative bias introduced by this operation is
of the order of $\alpha^2$~\cite{Zhao00,Necas20}, i.e.\ $<10^{-5}$ and thus
negligible.

\begin{figure}
\includegraphics[width=\hsize]{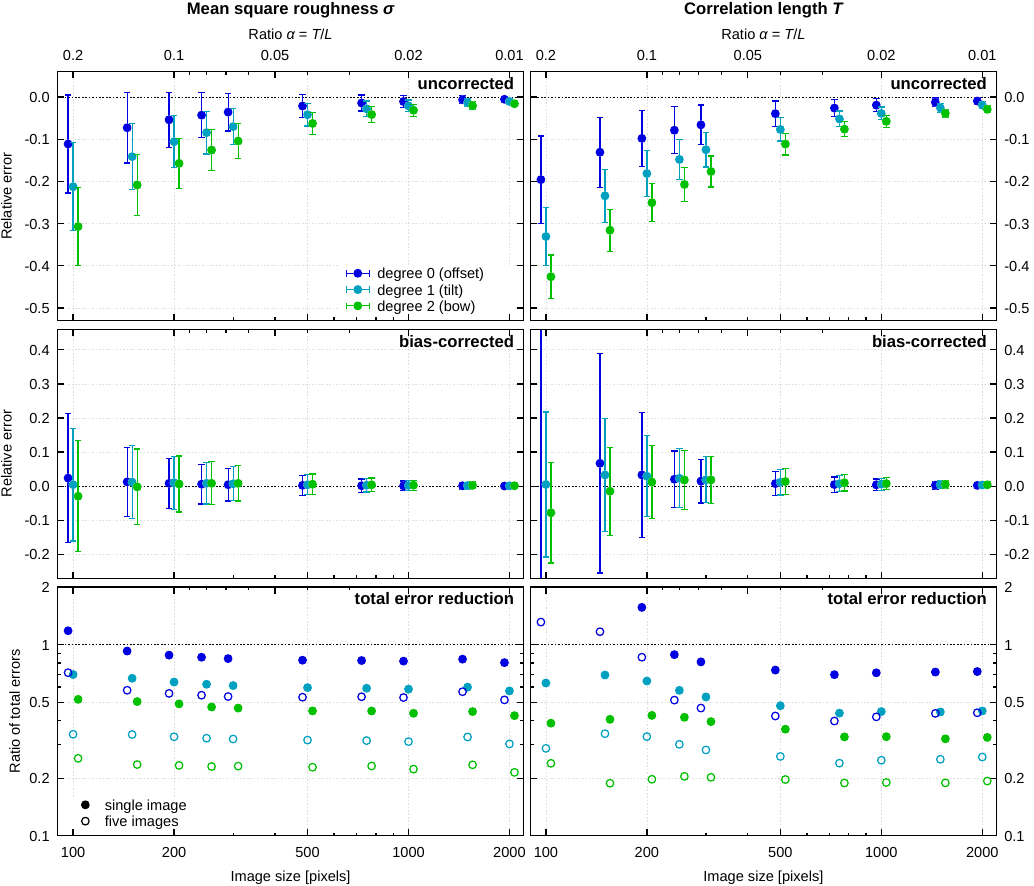}
\caption{Comparison of roughness parameters $\sigma$ and $T$ obtained from
uncorrected and model-free corrected ACF curves for a random pyramidal surface.
Error bars represent single-image standard deviations.
Results for different polynomial degrees are slightly offset horizontally for
visual clarity.}
\label{fig:pyramids-sigma-T}
\end{figure}

The results are plotted in \fref{fig:pyramids-sigma-T}. The overall trends are
similar as for the modified Gaussian model fitting. Conclusions
concerning polynomial degree and multi-image analysis remain unchanged. The
dependency on image size (or $\alpha$) is flatter, especially for $\sigma$.
Furthermore, $T$ is not improved at all for tiny images and degree 0. Unlike
for the explicit model, the correction in fact decreases the accuracy in this
case. It must be, however, emphasised that $T/L$ ratios around $0.1$ or larger
are not recommended, whether with correction or without.

We also tested how the correction depends on the ACF cut-off point by choosing
the interval from 10\% shorter to 40\% longer than to the first zero crossing.
The effect can be assessed using the accuracy of $\sigma$ and~$T$ or
differences between the corrected and true ACF curves. All the dependencies
are generally quite flat and often without any clear trend. This is
a reassuring result because it means the correction is not sensitive to the
cut-off point precise location. As expected, for very tiny images (large
$\alpha$) shortening the interval improves the accuracy somewhat. For large
images (small $\alpha$) the trend was sometimes slightly opposite.
Overall, however, cutting at the first zero crossing or moderately beyond it
appeared to work well.

\subsection{Rough thin film---inversion}

A test with real rough surface would ideally be done with a sample whose
ACF is precisely known. However, even standard rough samples do not have the
ACF specified. Furthermore, the objective is to verify that the bias caused by
limited area can be corrected. Meaning the resulting ACF is close to ACF which
would be obtained by measuring a very large (or infinite) area. The same
approach as in the previous section can thus be used. In fact, comparing
measurements on small and huge areas allows us to study the effect in
isolation---as opposed to comparison with a reference ACF where any observed
difference could have a variety of possible causes.

\begin{figure}
\includegraphics[width=\hsize]{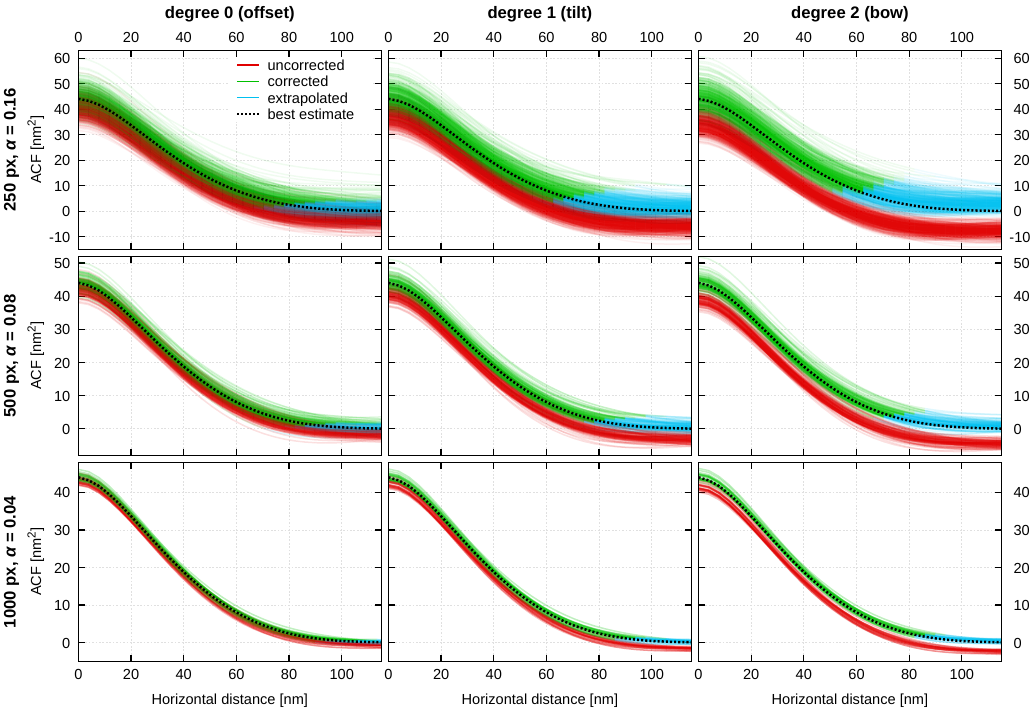}
\caption{Autocorrelation functions obtained by model-free correction
for rough SrO film surface, compared to uncorrected ACF and the best-estimate
ACF. Each curve corresponds to one subimage cut from the large base image.}
\label{fig:blind-SrO}
\end{figure}

An SrO thin film, prepared by atomic layer deposition, with large-scale
uniformly and isotropically rough upper surface was chosen for the
demonstration (see \fref{fig:surfaces}). The texture is formed by nanocrystals
and is clearly non-Gaussian. Images were acquired using a Bruker Dimension
Icon atomic force microscope in ScanAsyst mode with a standard ScanAsyst-air
probe and scan rate of 0.2\,Hz. In order to follow the 2D ACF route, a large
image without scan line artefacts is necessary. The absence of scan line
artefacts means 2D polynomial levelling is sufficient, leaving only bias
proportional to $\alpha^2$ (or higher powers). A scan of area
$12\times12\,\hbox{\textmu m}^2$ with pixel resolution of $3072\times3072$ was
selected for the evaluation. The correlation length to scan size ratio was
estimated as $\alpha\approx0.0037$, meaning the relative bias following from
background subtraction was $<10^{-3}$. The long scanning time resulted to
drift, which was estimated from the acquired image using Gwyddion {\it
Compensate drift} function. Its primary effect on the ACF is slight smearing
along the abscissa as distances in the $xy$ plane are distorted, in particular
in the slow scanning axis. The relative changes were estimated below
$1.5\times10^{-3}$ and thus negligible. The resulting ACF is plotted in
\fref{fig:blind-SrO} (each subplot) and separately also in \fref{fig:SrO-ACF}.

The large image was then cut to smaller images of various sizes and processed
as above, assuming subimages are reasonable approximations of measurements on
smaller areas. The uncorrected (red) and corrected (green) ACF computed for
each subimage are plotted in \fref{fig:blind-SrO} for three selected sizes and
all three polynomial degree 0--2. The corrected ACF curves were extrapolated
beyond the cut-off points by a simple subtraction of the last computed
correction from all further data (cyan).

The correction is clearly effective. The green (corrected) curves, although
spread slightly more than the red (uncorrected), are centred on the best
estimate ACF. Deviations are noticeable only for the highest degree and the
far ends of the curves, where there is a tendency to overcorrection.

\section{Discussion}
\label{sec:discussion}

We first remark on the normalisation factor in \eref{Gdiscrete} which is
sometimes taken to be $1/N$ instead of $1/(N-k)$ because of positive
definiteness and/or variance~\cite{Anderson71,Panda2016,Bittani2019}. It
corresponds to dividing the integrals in \sref{sec:biascalc} by $L$ instead of
$L-\tau$. However, the estimator with $N-k$ denominator is unbiased, or at
least it would be without background subtraction. Furthermore, a constant
denominator $N$ does not generalise to irregular regions and other cases where
varying amount of data is available for different distances
$\tau$~\cite{Necas12}. Therefore, in this context $N-k$ is the appropriate
choice.

\subsection{Interpretation of results}

\Fref{fig:blind-SrO} almost looks too good to be true. One has to be careful
with its interpretation. Everything was computed from the same large base
image. The clustering of the green curves around the best estimate shows that
we removed the bias tied to smaller scan areas. However, they do not
necessarily cluster around the \emph{true} ACF. In the example with synthetic
Gaussian and pyramidal data, the surfaces were uniform and could be made
infinite for all practical purposes. But for real rough surfaces, the issues
of uniformity, representativeness and the statistical character of roughness
are much more tangled. It should also be noted that roughness measurement is
also affected by tip sharpness and probe-sample interaction in
general~\cite{Sedin2001,Jacobs2017,GonzalezMartinez2012,Leach2014}, sampling
step~\cite{Jacobs2017,Sanner2022},
calibration, scanning speed and feedback loop
settings~\cite{GonzalezMartinez2012}, defects, and other
effects not analysed here as we attempt to isolate those related to the finite
area.

\begin{figure}
\includegraphics[width=0.5\hsize]{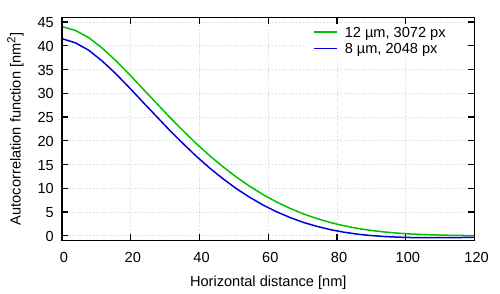}
\caption{Comparison of best-estimate ACF obtained using independent scans of
two different areas.}
\label{fig:SrO-ACF}
\end{figure}

The measurement of a neighbour region (somewhat smaller,
$8\times8\,\hbox{\textmu m}^2$) results in a slightly different ACF, as
illustrated in \fref{fig:SrO-ACF}. Subimages taken from this scan yield curves
centred on its own best-estimate ACF. The bias estimates for the two images
are approximately $3\times10^{-4}$ and $6\times10^{-4}$. The relative standard
deviations of $G(0)=\sigma^2$ are proportional to $\alpha$~\cite{Zhao00} were
estimated as $2\times10^{-3}$ and $3\times10^{-3}$. They are all too small to
explain the difference of almost 6\,\% between the two curves. The correlation
length does not capture the scale at which real textures can be considered
uniform. Although surface heights become uncorrelated for points considerably
farther apart than $T$, the texture itself varies along the surface. The
characteristic scale of these variations can be much longer than~$T$ even if
the texture is ultimately large-scale uniform. Scanning such large areas is
seldom feasible and we have to rely on multiple independent scans.

\subsection{Comparison with spectral density}

Two other functions are commonly used to characterise spatial properties of
roughness, height-height correlation function~\cite{Zhao00} (sometimes
also called structure function) and power spectrum density function (PSDF).
Height-height correlation function $H$ is directly related to ACF by
$H(\tau)+2G(\tau)=2\sigma^2$, so the results can be translated. PSDF is the
Fourier transform of ACF and is probably the most commonly utilised
function~\cite{Jacobs2017,Rutigliani2018}. The effect of levelling is
suppression of low-frequency components~\cite{Necas20_SR}.

The low-frequency components can be excluded from fitting, similarly how the
ends of spectral range are avoided in PSDF
stitching~\cite{Duparre2002,Gong2016,Panda2016,Jacobs2017,Klapetek17}. However, the peak
around zero frequency is where almost all the spectral weight lies. It is also
the least affected by noise, discontinuities and smoothing effects such as tip
convolution~\cite{GonzalezMartinez2012,Jacobs2017,Rutigliani2018}. It is often
critical in roughness analysis. However, it is the region worst affected by
levelling, and possibly in a non-trivial manner. In the case of ACF the worst
affected region is far from the origin and it is never used for analysis.
Around the origin, levelling manifests as the subtraction of a slowly varying
function. An approach similar to the one developed here can perhaps be
formulated also for PSDF---Ref.\ \citenum{Necas20}, for instance, gives hints
at spectral reinterpretation. However, what would be the equivalent of
model-free correction for PSDF is not clear.

\subsection{Zero crossing}

The model-free correction procedure relies on the true ACF monotonically and
quickly decaying to zero. In particular, sums of discrete ACF values must give
good approximations of integrals \eref{M0M1} (or similar integrals, but up to
$L$ instead of infinity). Even though it is true for many types of real
roughness, at least approximately, some violate this condition. For instance
if the surface is locally periodic/corrugated the true ACF crosses zero,
possibly many times. It may be possible to modify the correction procedure for
this case, but likely at the cost of reliability. And although the approach of
fitting $\hat G_k$ with biased model remains intact in principle, the first
zero crossing may no longer be a good choice of fitting cut-off.

All the procedures utilise the zero crossing for choosing the cut-off in some
manner. Must there always be a zero crossing? By splitting the sum $z_mz_k$
over all $m$ and $k$ into triangular parts and correcting for the
double-counted diagonal
\begin{equation}
\sum_{m,k=0}^{N-1} z_m z_k
= 2\sum_{k=0}^{N-1} \sum_{m=0}^{N-1-k} z_{m+k} z_k - \sum_{k=0}^{N-1} z_k^2
= 2\sum_{k=0}^{N-1} (N-k)\hat G_k - N\hat G_0\;.
\end{equation}
The left hand side is zero since the mean value of $z$ is zero. Therefore,
\begin{equation}
-\frac N2 \hat G_0 + \sum_{k=0}^{N-1} (N-k)\hat G_k
= \frac N2 \hat G_0 + \sum_{k=1}^{N-1} (N-k)\hat G_k
= 0
\end{equation}
and $\hat G_k$ must take both signs. As for the crossing location, the leading
term approximation of the analytical models or \eref{GAGc} is
a small constant (proportional to $M_0[G]$ or $S_0[G]$). If ACF decays
quickly, the first zero crossing occurs when the true ACF is equal to this
constant. And this is also when $S_0[G]$ and $S_1[G]$ can be assumed to give
good approximations to the corresponding integrals.

For biased model fitting, the heuristic zero-crossing rule is further
supported by the following:
\begin{itemize}
\item The rule is simple and easy to implement both manually and in code.
\item Fitting only data of the ACF apex at origin is an ill-conditioned problem.
      The optimal bias--variance trade-off invariably includes the side slopes
      in the fit. Shortening the interval too much cannot be beneficial.
\item Although fitting beyond the zero crossing may be beneficial, often the
      ACF is not converged in this region and telling where useful data end
      is difficult.
\item Numerical simulations support the zero crossing as a good choice
      (\sref{sec:pyramids}).
\end{itemize}
Choosing the cut-off based on zero crossing for each data contributes to
the increased variance of bias-corrected results. When multiple ACF curves are
evaluated it may be preferable to choose a single cut-off based on all the
data and use it for all curves.

\section{Conclusion}
\label{sec:conclusion}

The goal of this work was to correct the finite-area bias in autocorrelation
function (ACF) evaluation in roughness measurements, which includes the
correction of parameters like the mean square roughness and correlation
length. Starting from the observation that it should be possible to express
the bias of measured ACF in terms of ACF itself, we developed
a self-consistent formulation and used it to propose two types of bias
correction. One was a modification of standard analytical ACF models to take
into account the bias of the data they should fit. The other was a model-free
correction procedure based on inverting the self-consistent relations by
solving a set of linear equations. Their effectiveness was tested using
simulated and measured data.

The two corrections behave similarly. They appear most helpful in the cases
when they are also needed the most, that is the common moderate scan line
lengths, as data for too short scan lines are not salvageable and for very
long profiles the bias may already be small. Furthermore, they are more
beneficial for higher levelling polynomial degrees for which the bias is
worse. Both also trade bias for variance and thus the accuracy improvement is
larger when multiple scans are evaluated.
Modified (biased) analytical ACF models do not require any fundamental changes
to the evaluation and can even be used to re-analyse existing raw ACF data.
Based on numerical results, the measurement of Gaussian roughness, fitting the
experimental ACF with a modified model has substantial advantages and few
downsides and can probably be recommended quite universally.
The model-free correction (inversion) procedure proposed for ACF of an unknown
form is computationally efficient and worked surprisingly well in the selected
test cases. A simple zero crossing based criterion was proposed for choosing
the subset of discrete ACF data to use in the inversion. However, open
question remains regarding the application of the procedure to ACFs of more
complicated forms as the simple criterion may then no longer be suitable. The
second correction method thus should be currently considered more an
interesting concept to explore in further works.

\subsection*{Acknowledgements}

I would like to thank my colleagues Marek Eli\'a\v{s} and Lenka
Zaj\'\i\v{c}kov\'a for the rough samples used in the experimental part.
This research was funded by Czech Science Foundation under project
GACR 21-12132J.
The CzechNanoLab project LM2023051 funded by MEYS CR is acknowledged for the
financial support of the measurements and sample fabrication at CEITEC Nano
Research Infrastructure.

\subsection*{References}
\bibliographystyle{iopart-num}
\bibliography{article}

\end{document}